# An Accelerated Method for Message Propagation in Blockchain Networks

Wei Bi[*], Huawei Yang, Maolin Zheng

Seele Tech Corporation, San Francisco, USA
weibi@seelenet.com

**Abstract:** Blockchain is based on a P2P network, supporting decentralized consensus of current cryptocurrencies. Since bitcoin and altcoins all utilize an underlying blockchain, they are therefore greatly affected by the performance of the P2P network. In bitcoin, the miners are engaged in a time-sensitive competition to solve a Proof-of-Work problem to extend the blockchain. This consequently raises a critical problem to minimize the time between the propagation of a winning block and the beginning of the next round of the competition. This paper proposes a method that selects a node's closest neighbors to make messages propagate in the whole network in time. The method measures the distance from a node to its neighbor using transmission latency; thus, the lower the latency, the closer the neighbor. Simulations showed a good rate of decrease in average propagation latency and maximum propagation latency, compared to the classic method. Furthermore, this paper not only proposes the principle of establishing connections based on latency, but also evaluates the influence of the number of simultaneously established connections.

## 1    Introduction

A blockchain is a digitized, decentralized, public ledger of all cryptocurrency transactions. It allows market participants to keep track of digital currency transactions without central recordkeeping. Originally developed as the accounting method for the virtual currency bitcoin, blockchains, which use what's known as a distributed ledger technology, are appearing in a variety of commercial applications today.

A blockchain network is a decentralized peer-to-peer (P2P) network that transports all data needed to support the cryptocurrency system. The main goals of such a network are, firstly, to allow members of the network to synchronize their view of the system state, and secondly, to disseminate peer information in order to allow peers to reenter the system after a disconnection [1].

The bitcoin system is one of the most popular cryptocurrencies running on a blockchain network. It needs to disseminate different kinds of information, including transactions and blocks [2]. In bitcoin's P2P network, nodes randomly connect to other nodes. Transactions and blocks are transmitted over this network via these nodes until each has received all required data, which is inefficient [3]. This P2P network can be relatively slow. As such, miners (and pools) sometimes waste hash power mining on top of an old block while a newer block is finding its way through the network. Transmission delay, therefore, benefits pooled mining as well as geographic clustering of miners, incentivizing a more centralized mining topology. This is generally considered one of the bottlenecks for scalability because larger blocks (which can include more transactions) would propagate even slower [4,5].

Message propagation in bitcoin is shown in Fig.1. Each node receives the transaction request message, updates its own copy of the ledger, and passes on the message to nearby nodes [6]. So, a message will travel a long path to reach a far node situated at the corner of the network.

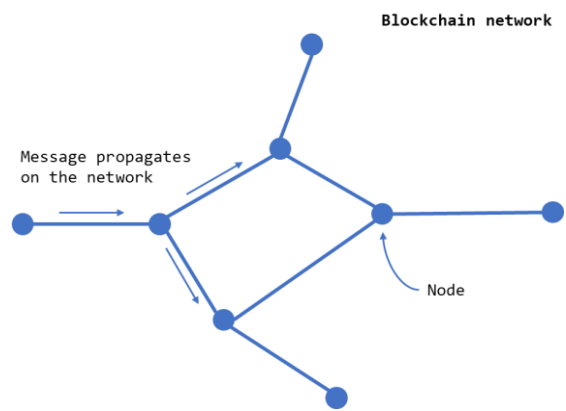

**Fig.1** Message Propagation via Neighbors in Bitcoin

The P2P system aims to share information among a large number of users without the assistance of explicit servers. Peers form an overlay topology which might be far different from the underlying physical network topology. The underlying blockchain network's propagation time is critical, especially for miners, because every millisecond of delay increases the chances that another block, found at about the same time, wins the "block race." However, the core nature of the blockchain is not beneficial for this: P2P networks are relatively slow, the overlay cannot perceive the underlying network, and nodes connect to neighbors in a random manner [7].

This paper proposes an optimized method to improve the speed of message propagation and is described in the following sections. Included are related researches in Section 2, the new method proposed in Section 3, method evaluation experiments in Section 4, and the actual and potential influences in the last section.

## 2  Related Research

Over the past years, several projects have been in development to increase the speed of block propagation. These projects focus on roughly two main issues: block compression to limit the amount of data that needs to be propagated over the network and relay speeds to cut the time it takes for blocks to propagate.

Compact Blocks, developed by bitcoin Core developer Matt Corallo, is a trick designed to decrease data-transmission. But if a node has not yet received the initial transaction before receiving the hashes, it cannot select the corresponding transaction. Additionally, in rare cases a wrong transaction may hash into a right hash, fooling a node into believing it received the right transaction. Xtreme Thinblocks, an option included in bitcoin Unlimited, is similar to Compact Blocks in many ways. For example, rather than sending all transaction data, Xtreme Thinblocks transmit more compact hashes. Weak Blocks, a relatively old idea, are essentially "almost valid" blocks. They are normal blocks in every way: They include transactions and all the rest — except that the hash of its header doesn't start with enough zeroes [8].

The Fast Relay Network is a relatively straightforward relay network that has existed for some years, and its downside is that this network is relatively centralized. And while the software is open source, no one else has set up a similar open relay network so far. The Falcon relay network uses a technique called "cut-through routing," where nodes don't wait to receive all packages to forward them. Falcon has one disadvantage: nodes can only truly validate a full block after they have received all required packages. FIBRE is built on UDP which allows FIBRE to use a nifty trick known as Forward Error Correction (FEC). This lets nodes reconstruct all of the transmitted data even if some of it gets lost on the way [9].

In addition, other methods consider minimizing the payload of the message (e.g. transactions in the block) to speed up network propagation. The idea of an Invertible Bloom Lookup Tables (IBLT) [10,11] is to squash together (using hashing and XOR) all of the transaction data into a fixed-size data structure. This method is restricted in several conditions, canonical ordering of transactions in blocks, similar policy for selecting which mempool transactions go into blocks, and peer sending an IBLT large enough with transactions that are not yet in our mempool.

Early research also paid attention to underlying network protocols, thus making the conversation complete as soon as possible. QUIC protocol [12], which implements TCP-like properties at the application layer atop a UDP transport, solves a number of transport-layer and application-layer problems experienced by modern web applications. Although QUIC reduces connection and transport latency, it didn't have any special treatment for P2P.

The above research focuses on accelerating message dissemination in the form of message compression or a relay network. They ignore the fact that P2P network problems should be resolved internally by the network and not by outside parties. This paper optimizes and fully exploits the characteristics of the network itself, which is the original source of creativity.

## 3  CNS Method

In a blockchain network, when a new node boots up, it must discover other nodes in the network to participate. To start this process, a new node must discover at least one existing node on the network and connect to it. The geographic location of other nodes is irrelevant as the blockchain network topology is not geographically defined. Therefore, any existing bitcoin nodes can be selected at random.

Each node can make a connection to another one, independent of the underlying network. The behavior of the blockchain is typical because it is a P2P overlay network. In reality, packets on this connection will go through many network links, which form a route from the initial node to its peer. The attributes of the route (or path) subsequently affect the connection's efficiency.

In message transmission, the distance between the connected peers can be measured by latency. The round-trip-time (RTT) metric plays such a role in popular network measurement environments [13]. RTT has been widely used as a metric for peer/server selection in classic Internet applications, such as streaming, tree-based multicast services, and other UDP and TCP-based services. In fact, RTT is used as a solution to infer forward and reverse delays by many protocols, including blockchains based on P2P networks.

In this paper, RTT denotes the distance between two nodes. The smaller the value, the closer the nodes. In a P2P overlay network, any two nodes are directly connected, thus becoming neighbors. While participating in blockchain, a node constructs a list of neighbors as potential peers to connect to.

In a full-connection P2P network with 10 nodes, there are 90 (i.e., n(n - 1), where n is the number of nodes) bi-directional connections. For example, a node connects to several selected peers, thus resulting in the sparse topology shown in Fig.2.

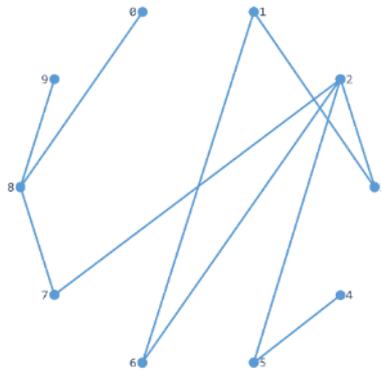

**Fig.2** A Connected P2P Network with at Least 1 Degree

The topology depicts a 10-node, 12-edge P2P network, where the nodes are indexed from 0 to 9. In this simple situation, each node has at least one connection (node degree is greater or equal to 1) to the network, thus making a connected graph. This connectivity allows messages from one node to eventually reach any node in the network. A connection between peers is depicted as a line and is internally associated with a value representing its transmission latency. The nodes, connections, and their latency are listed in Table 1.

**Table 1.** Nodes' Connections in a P2P Network and their Latency

| # | Connection | Latency(ms) |
|---|---|---|
| 0 | (0,6) | 1431 |
| 1 | (1,5) | 1252 |
| 2 | (2,4) | 1258 |
| 3 | (3,4) | 948 |
| 4 | (4,2) | 1258 |
| 5 | (4,3) | 948 |
| 6 | (5,1) | 1252 |
| 7 | (5,8) | 1471 |
| 8 | (6,0) | 1431 |
| 9 | (7,9) | 1445 |
| 10 | (8,5) | 1471 |
| 11 | (9,7) | 1445 |

A node must maintain a list of neighbors when it's a member of a blockchain network. The classic and simplified procedure when this happens is shown in several steps following.

List of RNS method steps

> Step 1: Initiate neighbors list when a node first connects to or reconnects to the network.
> Step 2: Refresh its neighbors list while the node finds a new neighbor or a lost neighbor.
> Step 3: Send messages via selected neighbors to the rest of the blockchain network.

The procedure consists of three steps, none of which apply any special rules. From this perspective, the procedure behaves in a free or random manner. We call this way for selecting neighbors the "Random Neighbors Selecting" (RNS) method. Although RNS works well in current deployed blockchain implementations, such as bitcoin and Ethereum, the network suffers from delayed messages.

Therefore, we propose a new method to optimize the transmission latency in blockchain networks, selecting the closest neighbors and using them to spread messages out. This Closest Neighbors Selecting (CNS) method is described in several steps following.

List of CNS method steps

>Step 1: Initiate a potential neighbors list, where each neighbor is associated with an infinite latency value.
>Step 2: Refresh the list of neighbors when new nodes are found or old nodes are lost, and refresh each neighbor's latency value via RTT measurement.
>Step 3: Sort the neighbors in the list by its latency value, with the smaller values first.
>Step 4: Send messages by selecting neighbors in order in the list to nodes all over the blockchain.

From these descriptions of the two methods above, some slight but important changes can be seen in CNS. Firstly, each node is associated with a metric of connection latency to other nodes, indicating distance between the paired two nodes. Secondly, the latency value for each node is measured via message sending and echoing, representing a point-to-point RTT. Thirdly, the neighbors are selected in a priority order, rather than a random way of equal opportunity.

The first change can be implemented in the software by using a simple data list, which is refreshed as the blockchain network fluctuates. The third change can be dealt with a generic sorting algorithm, making the list sorted by RTT values from small to large. With these two easily solvable changes (included in steps 1, 3 and 4), we are left with the second change (included in step 2).

The second change focuses on a classic network measurement problem, including RTT measuring between a pair of nodes. RTT can often be measured as a by-product of the traffic flowing between two points [14]. Active measurement systems inject specific measurement packets. Passive measurement systems use either direct measurement packets or observe existing traffic.

We assume that both ends are interested in evaluating the RTT, with the active one as a client and the passive one a server [13]. Referencing the Fig.3, we define $t_{Sc}$ as the time instant when the probe request is sent by the client-end, $t_{Rs}$ the time instant when the server-end receives the probe request, $t_{Ss}$ the time instant when the server-end sends the probe response, and $t_{Rc}$ the time instant when the client receives the probe response. Note that the client and server clocks do not need to be synchronized, therefore $t_{Sc}$ and $t_{Rc}$ represent the times as measured by the client clock, while $t_{Rs}$ and $t_{Ss}$ represent the times as measured by the server clock.

The probe request messages include 3 parameters:

$t_{Sc}(k)$, $t_{Ss}(k_{prev})$, $\Delta t_C(k) = t_{Sc}(k) - t_{Rc}(k_{prev})$

where $t_{Ss}(k_{prev})$ and $t_{Rc}(k_{prev})$ represent the most recently received values for these state variable.

The probe response messages include 3 parameters:

$t_{Ss}(k)$, $t_{Sc}(k)$, $\Delta t_S(k) = t_{Ss}(k) - t_{Rs}(k)$

This way, both ends of the tunnel can evaluate the RTT delay from the probe packets without keeping state information, as follows:

On the client-end:

$RTT_c(k) = t_{Rc}(k) - t_{Sc}(k) - \Delta t_S(k)$.

On the server-end:

$$RTT_s(k) = t_{Rs}(k) - t_{Ss}(k_{prev}) - \Delta t_C(k).$$

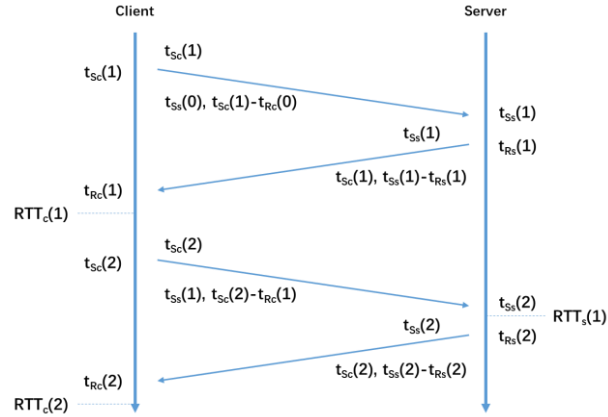

**Fig.3** Time Sequence for RTT Evaluation Procedure

So, with RTT values, we can select the closest neighbors for network-wide message propagation, acquiring better performance. Blockchain with low latency, therefore, creates a foundation for more efficient application delivery.

## 4      Experiments

Compared to the RNS method, CNS adds many new features to improve performance. This section describes the evaluation model and experiments for CNS.

The evaluation network is modeled as a connected graph. Formally, a graph is a pair of sets. This can be portrayed by G = (V,E), where V is the set of vertices and E is the set of edges. E is made up of unordered pairs of elements from V.

A DiGraph is also a pair of sets, D = (V,A). V is the set of vertices and A is the set of arcs. A is made up of ordered pairs of elements from V.

In the case of digraphs, there is a distinction between `(u,v)` and `(v,u)`. Usually the edges are called arcs to indicate the presence of direction.

In this section, for simplicity, the network is represented by a DiGraph, and with two direction edges having the same weight. P2P networks with 100 nodes and different degrees are shown below. The next 6 graphs (a, b, c, d, e, f) in Fig.4 are created with degrees (2, 3, 4, 5, 10 and 20), and lines randomly connect pairs of nodes. The number of nodes and the degree of the network play an important role in the following experiments.

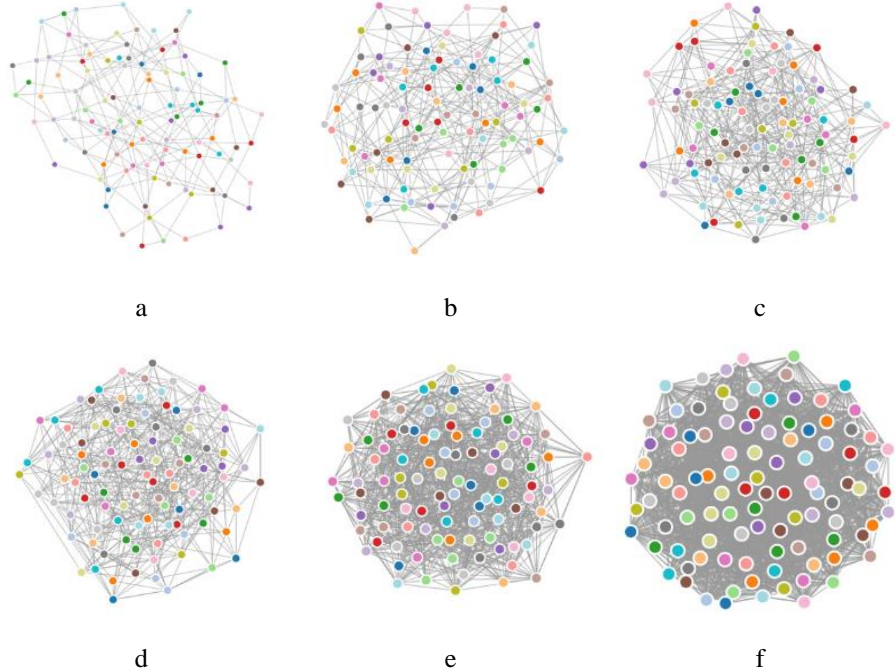

**Fig.4** Connected P2P Networks with Degree Increasing

For the CNS method, each connection is associated with a latency value, which is paired with a neighbor. The latency values of the whole network are assumed to follow a normal distribution (Equation 1).

$$f(x; \mu, \sigma) = \frac{1}{\sigma\sqrt{2\pi}} \exp\left(\frac{-1}{2}\left(\frac{x-\mu}{\sigma}\right)^2\right) \tag{1}$$

We set μ=2000, mean connection latency to 2000ms, and standard deviation σ=500 in this paper.

**Experiment 1**

This experiment shows nodes' average latency in a 40-node network, where each node has 5 closest neighbors (directional). The node's average latency is the average of the latency values from the node to its 5 neighbors. In Fig.5, the two solid lines are the nodes' average latency in RNS and CNS methods separately, with averages of 2006.19ms and 1194.41ms. The dashed line indicates the percentage decrease of CNS relative to the RNS, with an average of nearly 40%.

In the same experiment, we also compute the maximum of the 5 values representing the node's latency to its neighbors. The CNS average is 1381.6ms, while RNS is 2605.9ms, presenting a decrease of almost 50% (Fig.6).

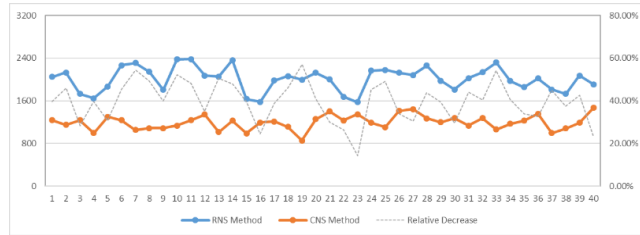

**Fig.5** 5 Neighbors' Average Latency in a 40-node Simulation Network

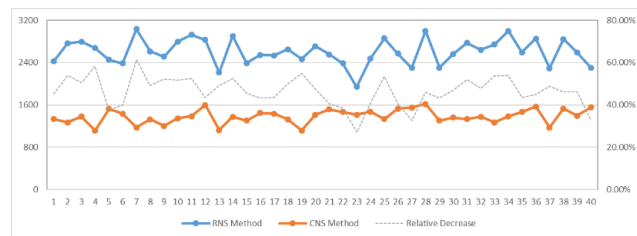

**Fig.6** 5 Neighbors' Maximum Latency in a 40-node Simulation Network

The RNS method selects 5 random neighbors, and the CNS method selects 5 closest neighbors, which accounts for the different results above. Obviously, CNS has an advantage over the RNS approach.

**Experiment 2**

This experiment compares results from the perspective of the whole network, different from the previous node of view. Each point in the curves below is a summary of a network, which is created with a different network degree (the number of connections to a node) each time. As the degree of the network increases, the CNS method's advantage of average network latency is gradually reduced (Fig.7).

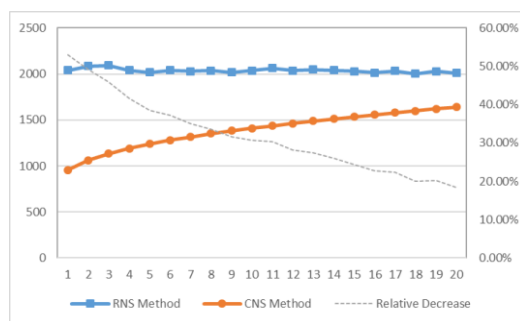

**Fig.7** Network's Average Latency Across 1-20 Degree

This experiment also computes a network's average value from each node's maximum latency. The curves in Fig.8 show similar results as above.

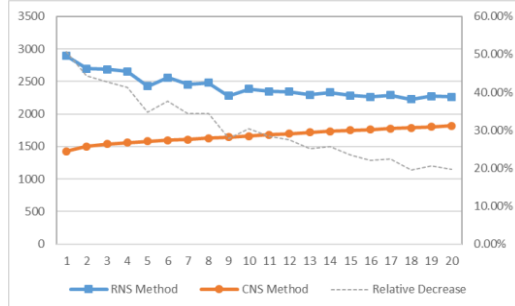

**Fig.8** Network's Maximum Latency Across 1-20 Degree

The RNS method and the CNS method can select more neighbors to share the traffic burden, but as the number of created connections increases, the difference between the two algorithms decreases. This indicates that at some critical point, the CNS method no longer has an advantage; this point is related to actual network situations.

## 5    Conclusion

Blockchains are increasingly becoming more popular in the world today. We continually find newer ways to leverage the power of the blockchain for intuitive applications that provide solutions to real-world problems. This paper is about the efficiency of blockchains [15] while supporting higher applications. The P2P latency metric was proposed, and experiments have shown that the method of selecting nearest neighbors (i.e. CNS) is conducive to faster message propagation.

CNS and these experiments make clear that by purposely selecting low-latency neighbor nodes, message propagation can be effectively improved. Also, message exchange acceleration can be implemented without third party (such as relay networks) services. This makes the following points a reality. Firstly, the new block miner should win the Proof-of-Work competition and be rewarded instead of being taken away by a successor. Secondly, the network greatly avoids useless power consumption with efficient message transmission, and thus, runs in an energy-efficient style [16]. Thirdly, fast message exchange minimizes the amount of memory and bandwidth used by the nodes on the network. Finally, this optimal latency model could promote the evolution of performance-awareness networks or drive an evolution of the underlying blockchain structure.

This paper proposed the principle of selecting closest neighbor nodes, and this is efficient blockchain information exchange. In practice, the node location weights are introduced in consideration of the decentralization problem, etc. Further research will develop the robustness and scalability of blockchain networks while optimizing their efficiency [17].